# Classification of infrastructure networks by neighborhood degree distribution

[1]Orazio Giustolisi, [1]Antonietta Simone, [2] Luca Ridolfi
[1]Politecnico di Bari, Bari, Italy. orazio.giustolisi@poliba.it - antonietta.simone@poliba.it
[2]Politecnico di Torino, Torino, Italy. luca.ridolfi@polito.it

**Abstract**

A common way of classifying network connectivity is the association of the nodal degree distribution to specific probability distribution models. During the last decades, researchers classified many networks using the Poisson or Pareto distributions. Urban infrastructures – like transportation (railways, roads, etc.) and distribution (gas, water, energy, etc.) systems – are peculiar networks strongly constrained by spatial characteristics of the environment where they are constructed. Consequently, the nodal degree of such networks spans very small ranges not allowing a reliable classification using the nodal degree distribution. In order to overcome this problem, we here (i) define the "neighborhood" degree, equal to the sum of the nodal degrees of the nearest topological neighbors, the adjacent nodes and (ii) propose to use "neighborhood" degree to classify infrastructure networks. Such "neighborhood" degree spans a wider range of degrees than the "standard" one allowing inferring the probabilistic model in a more reliable way, from a statistical standpoint. In order to test our proposal, we here analyze twenty-two real water distribution networks, built in different environments, demonstrating that the Poisson distribution generally models very well their "neighborhood" degree distributions. This result seems consistent with the less reliable classification achievable with the scarce information using the "standard" nodal degree distribution.

## 1. INTRODUCTION

The complex network theory (CNT) is becoming one of the most powerful and versatile tool to investigate, describe, and understand the world (Barabasi [1]). Networks allow studying and interpreting a huge number of physical, biological and social processes. The examples range from social relationships to neural connections, from multi-species ecological interactions to financial and economic exchanges. Although each network exhibits its own topological and structural peculiarities, apparently very different networks share amazing similar features (Albert and Barabasi [2], Buchanan [3]).

In the last decades, CNT had an unrestrained development and researchers proposed novel approaches, metrics and theories to explore and disentangle network features (e.g., Boccaletti et al. [4]; Newman [5]). Nevertheless, the nodal degree distribution remains a relevant concept to classify networks. The nodal degree distribution describes the probability distribution of the number of edges incident each node of the network. Several network features are associated to the shape of such distribution. Erdos and Rény [6] [7] were the first to study the nodal degree distribution of real networks introducing the *random* networks against *regular* networks. In the former case, the nodal degree distribution is random around an average value and the network is characterized by a high homogeneity; differently, regular networks have a constant degree of internal nodes and are characterized by an absolute homogeneity. The Poisson model is often used to describe nodal degree distribution of random networks, which are able to capture some features of real networks better than regular networks: e.g., being highly ordered, the shortest paths linking two nodes of regular networks are too large, while random networks show lower and more realistic values (e.g., Milgram [8]).

Later on, starting from regular networks, Watts and Strogatz [9] introduced the *small world* networks. They demonstrated the existence of the so-called small world effect, i.e. a behavior between the regular and random networks. The model by Watts and Strogatz [9] covers from regular networks to random networks using the Poisson distribution of nodal degrees and a probability of connections between two nodes: for small world networks, that probability is greater than zero of regular networks but lower than values typical of random networks. In the same period, Barabasi and Albert [10] proposed the *scale-free* networks in order to describe real networks characterized by non-homogeneous nodal degree distributions, where many nodes have a low degree and few nodes (called hubs) have a high degree. These distribution typically exhibits a Pareto (or power law) behavior.

The interest in classifying networks according to connectivity structures is related to capturing the emerging behaviors of real systems (e.g., Albert and Barabasi [2], Newman [5] [11]) For instance, the nodal degree distribution affects the network vulnerability to random failures and intentional threats. In fact, regular small world networks and random networks present a significant structural resistance to both random failures and intentional threats, while scale free networks show a very high structural resistance to random failures but a weak resistance to intentional threats (Albert et al. [12]). In this sense, the classification of real networks by associating their degree distribution to the Poisson or Pareto models is useful to assess network vulnerability.

In spite of the nodal degree distribution has a crucial role to assess the network behavior, for a relevant class of real systems this information is poorly informative; they are the so-called infrastructure networks. Remarkable examples are transportation networks (railways, roads, etc.), distribution networks (energy, water, gas, etc.), and urban street patterns. These networks are one of the most widespread and important examples of spatial networks, and they are characterized to be (i) located on a two-dimensional space and (ii) constrained by a number of spatial impediments (Barthelemy, [13]). This entails that the maximum nodal degree is generally very low, consequently the nodal degree distribution spans over a very limited range and, therefore, statistical inference results unreliable. It follows a quite paradoxical situation: on one hand, many theoretical and applicative works underline the key information embedded in the degree distribution; on the other hand, infrastructure networks have characteristics that make *standard* nodal degree distribution evaluation very elusive and difficult to interpret.

In order to overcome this problem, we introduce the *neighborhood* nodal degree spanning over a range of values wider than the *standard* nodal degree, resulting in a statistically more reliable classification of infrastructure networks (hereinafter we will specify as *standard* the usual nodal degree when a possible misunderstanding with *neighborhood* nodal degree can occur). The *neighborhood* nodal degree corresponds, for each node, to the sum of the degrees of nearest topological neighbors, namely of the adjacent nodes. We will demonstrate that the approach based on the *neighborhood* nodal degree makes more reliable the association of the connectivity structure of an infrastructure network to a Poisson or a Pareto distribution. We will show that the *neighborhood* nodal degree can be interpreted as a weighting of the network edges incident to each node by the degree of the nearest neighbors or adjacent nodes.

In order to show the effectiveness of the *neighborhood* nodal degree to classify spatial networks, we will focus on water distribution networks (WDNs). This is due to several reasons. Firstly, WDNs are planar urban infrastructure networks, strongly constrained by external geometrical/environmental factors – like the landscape topography, the structure of the cities (e.g., street network, building size, etc.) – decreasing the connectivity probability with the distance between two nodes. It follows that WDNs classification by the *standard* nodal degree distribution is very difficult and uncertain (Yazdani and Jeffrey [14]) because the very low values of the maximum *standard* nodal degree characterizing those networks (generally around five and quite always lower than ten). Secondly, and somewhat surprisingly, CNT literature reports very few investigations on WDNs in spite of their ubiquity, topological variety, and importance in everyday life.

Thirdly, the availability of real WDNs is quite rare due to the sensible data associated with water supply service. Thus, the unusual availability of detailed topological data of twenty-two real WDNs built in different urban environments, allows a robust and well representative databased investigation about the promising use of *neighborhood* nodal degree to classify real infrastructure networks.

In the following, we will demonstrate that the network connectivity structure of WDNs generally follows the Poisson distribution. The result is consistent with the fact that technicians generally design the network connectivity structure of WDNs using a criterion of redundancy beyond the hydraulic capacity requirements. This fact provides a low vulnerability of the connectivity structure with respect to any kind of failure event.

## 2. DEGREE DISTRIBUTION MODELS

The Empirical nodal degree distribution $P(k)$ is defined as the fraction of nodes in the network having degree $k$. Hence

$$P(k) = \frac{n_k}{n} \tag{1}$$

where $n_k$ is the number of nodes having degree $k$ and $n$ is the total number of nodes in the network. The formulation of the Poisson distribution, approximating the binomial one, for nodal degrees (Watts and Strogatz [9]) is

$$P(k) = \binom{n-1}{p} p^k (1-p)^{n-1-\langle k \rangle} \approx \frac{e^{-\langle k \rangle} \langle k \rangle^k}{k!} \tag{2}$$

where $p$ is the probability of connection between two nodes and $<k>$ is the average nodal degree of the network. The Poisson distribution models random networks and it exhibits a peak value at $<k>$ and a high probability of nodal degree around $<k>$, i.e. a high homogeneity of the degree distribution. The nodal degree distribution of small world networks has essentially the same features as the random networks (Barthelemy [13]). Therefore, the Poisson distribution can apply to small world networks arguing a narrow distribution of degrees, i.e. a very high homogeneity of the nodal degree distribution, which degenerates to a single degree (considering only internal nodes) for the case of regular networks, characterized by an absolute homogeneity. In fact, Watts-Strogatz [9] introduced the small world networks starting from a ring lattice regular network, increasing the probability of connection, $p$.

In the present study, we will show that a Poisson distribution models very well the *neighborhood* nodal degree distributions of WDNs allowing arguing that it also models the *standard* nodal degree. It is to note that Watts-Strogatz [9] assumed the probability $p$, determining the average degree of the network equal to

$$\langle k \rangle = p(n-1) \tag{3}$$

In the case of the infrastructure networks, the average nodal degree strongly depends on spatial constraints and, generally, it is very low. E.g., in the case of WDNs, $<k>$ ranges from 2 to 3. Hence, we can write

$$p = \frac{\langle k \rangle}{n-1} \approx \frac{2.5}{n-1} \quad \Rightarrow \quad p \propto n^{-1} \tag{4}$$

Namely, the probability of connection results inversely proportional to the size of the network because of the spatial constraints strongly limiting the range of variability for $<k>$. Consequently, the Watts-Strogatz [9] scheme has to be carefully used with respect to the probability $p$, because Eq. (4) shows that $p$ is generally very low for the connectivity structure typical of infrastructure networks and it decreases with the network size.

Finally, the formulation of the Pareto (or power law) model for the nodal degree distribution is

$$P(k) \approx k^{-\gamma} \tag{5}$$

where $\gamma$ is a constant generally ranging from 1.5 to 3 (Newman [5]; Barthelemy [13]). The model of Eq. (5) applies to the scale free networks, which are non-homogeneous with respect to the nodal degree distribution. In other words, few nodes with a large degree (hubs) characterize scale free networks.

## 3. NEIGHBORHOOD NODAL DEGREE DISTRIBUTION

The starting point of our reasoning is that a reliable classification of spatial networks by means the structure of the *standard* degree distribution (e.g., if it exhibits a Pareto or Poisson law behaviour) is generally hampered by the very low maximum nodal degree characterizing this class of networks. This fact is a consequence of the spatial limits constraining the network topology at local scale (i.e., at the node scale) (Barthelemy [13]; Lämmer et al., [15]). E.g., the topographic structure of a city typically involves the crossing of three-four roads and rarely the nodal degree of a node-square exceeds six-eight. It follows that infrastructure networks inheriting these urban constrains (e.g., water and gas distribution networks) exhibit a very narrow range of nodal degrees, which makes very difficult to model the degree distribution in a statistically meaningful way.

In order to classify infrastructure networks still using the degree distribution concept, we here introduce a more suitable idea: the *neighborhood* nodal degree. I.e., we define a degree distribution involving the nearest neighbors or adjacent nodes. The *neighborhood* degree distribution of each node is the sum of the *standard* degrees of the topologically nearest (i.e., adjacent) nodes. We propose the *neighborhood* nodal degree distribution to classify infrastructure networks. The approach allows significantly increasing the maximum nodal degree and, consequently, the range of values spanned by *neighborhood* nodal degrees. This fact makes much more reliable the classification, because a greater number of points supports the robust identification of a specific statistical distribution.

The formulation of the neighborhood degree is

$$k_n(i) = \sum_{j \in N(i)} A_{ij} \cdot k(j) \qquad (6)$$

where $k_n(i)$ is the "neighborhood" degree (involving adjacent nodes) of the *i*-th node, $A_{ij}$ are the elements of the adjacency matrix, $k(j)$ is the *standard* degree of the *j*-th node, and $N(i)$ is the topological neighborhood of *i*-th node, i.e. the set of adjacent nodes. Therefore, $k_n(j)$ is the product between the *standard* nodal degree and the *i*-th row of the adjacency matrix providing a non-null value for the nearest/adjacent nodes only.

Figure 1 shows an illustrative example of the neighbourhood degree evaluation. Consider the portion of a network reported in the Figure 1 and focus on the node 1, Figure 1(b). Its neighbourhood is constituted by nodes labelled {2,3,4}, whose nodal degrees are equal to $k(2)=4$, $k(3)=3$, and $k(4)=3$, respectively. In this case, the *neighbourhood* degree of node 1 results equal to $k_n(1)=4+3+3=10$, i.e. the sum of the degree of the three adjacent nodes. Similarly, the strategy assigns to the node 2 (see Figure 1(c)) a degree $k_n(2)=3+3+3+4=13$, i.e. the sum of the degree of the four adjacent nodes {1,3,5,6}.

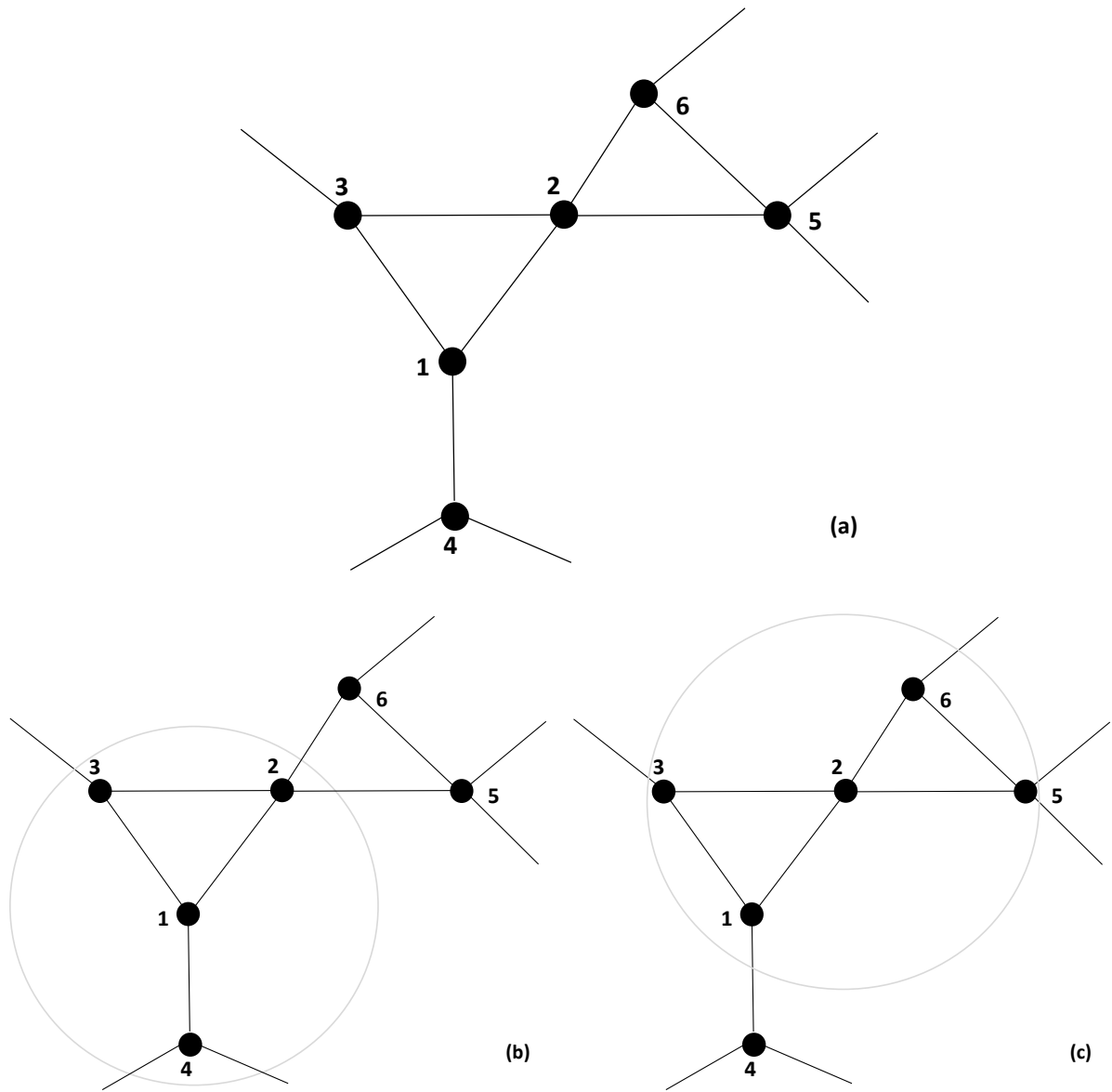

Figure 1. (a) Example of a portion of a network. Panels (b) and (c) show the *neighbourhood* of node 1 and node 2, respectively.

The *neighborhood* nodal degree can be interpreted as the node strength (Boccaletti et al. [4]) of a suitable oriented weighted network, which has the structure of the infrastructure network adjacency matrix. Each non-null cell $i,j$ has a value corresponding to the link weight between $i$ and $j$ which is assumed equal to degree of the starting node. E.g., the Figure 2(a) and (b) reports the weights to be assigned to the links incident node 1 and 2 in order to interpret the proposed strategy as a weighting of the adjacent matrix. In the Figure 2(a) the link 1-2 has a weight, $w_{12}$, equal to 3 considering the direction from node 1 to node 2 because $k(1)=3$, while the link 2-1 has a weight $w_{21}=4$ (considering the direction from node 2 to node 1) being $k(2)=4$; similarly for the links 1-3, 3-1, 1-4, and 4-1. The strategy is similarly applied to node 2.

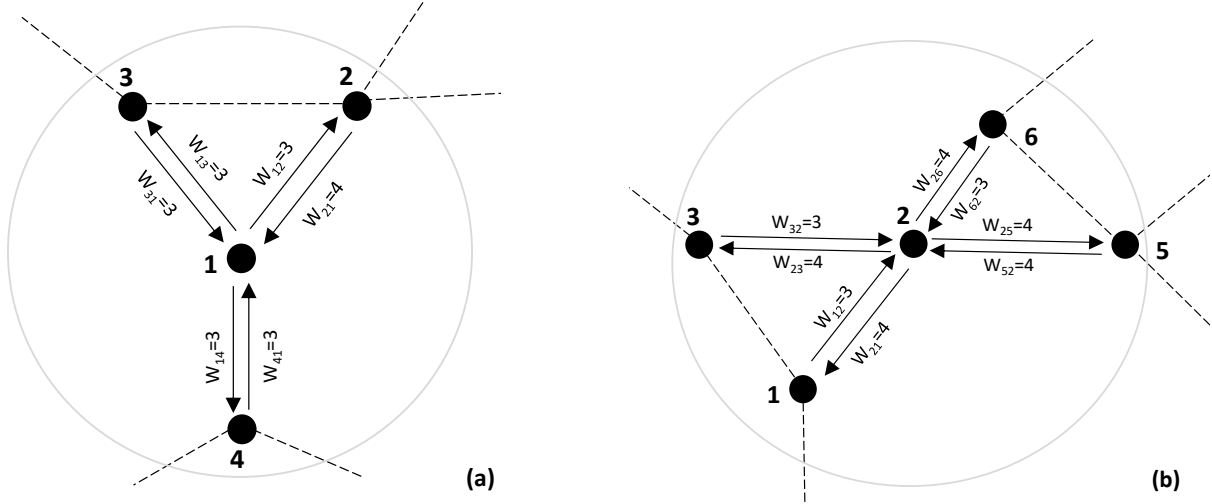

Figure 2. Weights assigned to links incident nodes 1 (a) and 2 (b).

Introducing such oriented weighted network, the neighbourhood degree can be seen as the node strength (Boccaletti et al. [4]), namely

$$k_n(i) = \sum_{j \in N(i)} w_{ji} \qquad (7)$$

where $w_{ij}$ are the previously defined topological weights. In this way, the weights can be interpreted as a sort of flow of topological information, which travels on each link, and the *neighbourhood* degree is the overall information that flows into each node from the nearest neighbours.

## 4. CLASSIFYING THE NETWORK STRUCTURE: NEIGHBORHOOD VERSUS STANDARD NODAL DEGREE DISTRIBUTION

In order to test the effectiveness of *neighbourhood* nodal degree to classify infrastructure networks, we start showing some results about a small but realistic infrastructure network (see Figure 3) while in the next section results about a significant number of real networks will be described. Figure 3 shows the network layout of the so-called BBLAWN network (Giustolisi et al. [16]). The network is composed of 390 nodes and 439 links and it is frequently used as benchmark network in studies about WDNs (Ostfeld et al., [17]). Figure 4 reports the Empirical density distributions, $P(k)$ and $P(k_n)$, and the corresponding cumulative distributions, $P_{cum}(k)$ and $P_{cum}(k_n)$, of the BBLAWN related both to the *standard* nodal degree, $k$ (Figures 4(a, c)), and the *neighbourhood* nodal degree, $k_n$ (Figures 4(b, d)). The same figure reports the corresponding theoretical Poisson and Pareto distributions, in order to compare them to the Empirical ones. In order to allow an easier comparison in the case of Pareto distribution, Figures 4(e, f) report the cumulative distributions also in logarithmic scale.

The diagrams (a), (c) and (e) of Figure 4 reveal that the identification of a specific probabilistic model is very difficult when using the *standard* nodal degree: the maximum degree is five and, consequently, only five points are available to fit the distribution. It follows that to discern between the Poisson and Pareto distributions is difficult and statistically unreliable. The only reasonable conclusion arguable from the diagrams (a), (c) and (e) of Figure 4 is that the *standard* nodal degree distribution seems to be qualitatively more similar to the Poisson model than the Pareto one.

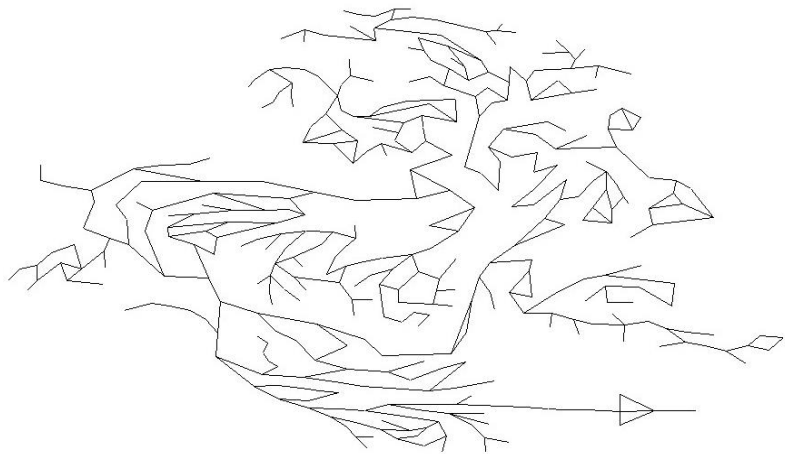

Figure 3. BBLAWN layout.

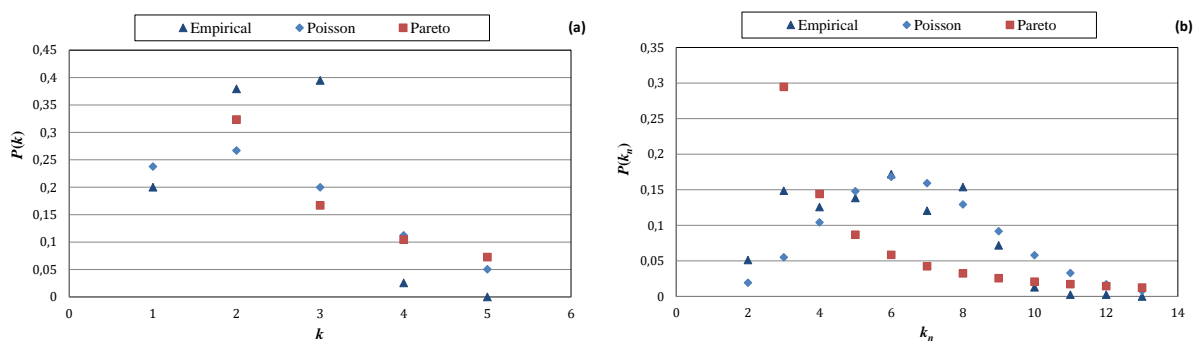

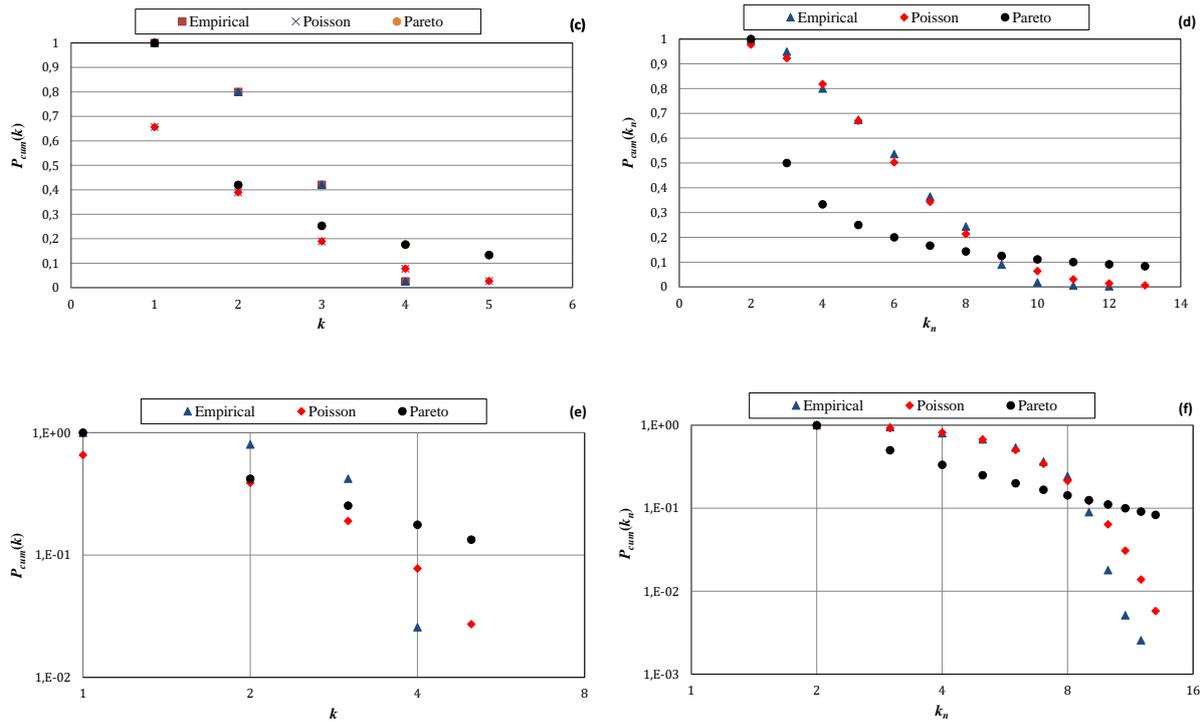

Figure 4. Empirical *standard* and *neighbourhood* nodal degree versus Poisson and Pareto distributions.

On the contrary, diagrams (b), (d) and (f) on the right side of Figure 4 (related to the *neighbourhood* nodal degree) allows discerning between the Poisson and Pareto distributions being the maximum degree equal to 12 and, therefore, a wider interval of degrees is now available. In fact, the eleven points (degrees) make reliable the statistical analysis as will be reported in Table I. About the patterns of the diagrams in the right side of Figure 4, it is evident that the Empirical distributions follow very well the Poisson distributions especially referring to the cumulative ones. The different pattern between Empirical data and the Pareto distribution is more evident in the logarithmic scale of the cumulative distribution.

We systematically compared *standard* and *neighbourhood* nodal degree distributions for twenty-two networks corresponding to existing WDNs. Table I reports their relevant characteristics. In particular, it is worth noting that network size spans two order of magnitude; in fact, the number of nodes ranges from 390 to 26.761, while the number of links varies from 439 to 32.096. In spite of the size variability, the average nodal degree remains always very low – ranging from 2.08 to 2.59 – as well as the maximum nodal degree, ranging from 5 to 11. These values confirm the situation reported in the previous paragraphs about the BBLAWN. The small number of points of the Empirical distribution frustrates any tentative to infer a statistical model. In contrast, the *neighbourhood* nodal degree depicts a very different picture: the average and maximum values range from 4.68 to 7.98 and from 12 to 33, respectively. The minimum *neighbourhood* nodal degree results equal to two as it corresponds to the extreme nodes of network branches composed by serial pipes/links. Furthermore, in logarithmic diagrams, Empirical curves does not report any final value. This is because the log of zero does not admit finite value.

Table I. Relevant data of the 22 networks (for few benchmark real WDNs the inhabitants are not available)

| WDN Name | Node # | Pipe # | Mean Standard | Min Standard | Max Standard | Mean nearest | Min nearest Neighbor | Max nearest Neighbor | KS test [%] | Length [Km] | Inhabitants [x1000] |
|---|---|---|---|---|---|---|---|---|---|---|---|
| BBLAWN | 390 | 439 | 2.25 | 1 | 5 | 5.69 | 2 | 13 | 78.6 | 57 | - |
| Big Town | 26,761 | 32,096 | 2.40 | 1 | 8 | 6.18 | 2 | 17 | 91.2 | 2,054 | 1,347 |
| Apulia 1 | 18,718 | 19,990 | 2.14 | 1 | 7 | 5.01 | 2 | 16 | 99.8 | 678 | 326 |
| BWSN | 12,518 | 14,314 | 2.29 | 1 | 6 | 5.80 | 2 | 17 | 99.9 | 1,844 | - |
| Apulia 2 | 5,288 | 6,116 | 2.31 | 1 | 6 | 6.01 | 2 | 17 | 99.9 | 277 | 151 |
| Apulia 3 | 5,036 | 5,848 | 2.32 | 1 | 6 | 5.98 | 2 | 16 | 99.8 | 199 | 107 |
| Norway 1 | 5,035 | 5,292 | 2.10 | 1 | 6 | 4.68 | 2 | 13 | 78.6 | 239 | - |
| Apulia 4 | 4,242 | 4,940 | 2.33 | 1 | 6 | 6.06 | 2 | 17 | 100.0 | 161 | 70 |
| Apulia 5 | 4,188 | 4,727 | 2.26 | 1 | 5 | 5.82 | 2 | 14 | 99.5 | 261 | 94 |
| Apulia 6 | 3,547 | 3,881 | 2.19 | 1 | 5 | 5.37 | 2 | 17 | 100.0 | 113 | 33 |
| Apulia 7 | 3,000 | 3,189 | 2.13 | 1 | 5 | 5.12 | 2 | 13 | 99.1 | 91 | 49 |
| Apulia 8 | 2,968 | 3,400 | 2.29 | 1 | 5 | 5.89 | 2 | 16 | 100.0 | 130 | 60 |
| Apulia 9 | 2,895 | 3,333 | 2.30 | 1 | 5 | 5.86 | 2 | 16 | 99.8 | 85 | 30 |
| Apulia 10 | 2,810 | 3,307 | 2.35 | 1 | 5 | 6.26 | 2 | 16 | 99.8 | 175 | 398 |
| Piedmont 1 | 2,784 | 2,894 | 2.08 | 1 | 5 | 4.89 | 2 | 12 | 98.5 | 171 | 49 |
| Norway 2 | 2,520 | 2,651 | 2.10 | 1 | 7 | 4.87 | 2 | 18 | 99.8 | 129 | 24 |
| Apulia 11 | 2,403 | 2,820 | 2.35 | 1 | 5 | 6.11 | 2 | 15 | 99.7 | 145 | 94 |
| Apulia 12 | 1,918 | 2,153 | 2.25 | 1 | 6 | 5.74 | 2 | 16 | 100.0 | 108 | 56 |
| Exnet | 1,776 | 2,300 | 2.59 | 1 | 11 | 7.98 | 2 | 33 | 25.8 | 594 | - |
| Apulia 13 | 1,762 | 2,098 | 2.38 | 1 | 5 | 6.38 | 2 | 17 | 100.0 | 70 | 25 |
| Apulia 14 | 1,263 | 1,428 | 2.26 | 1 | 5 | 5.78 | 2 | 15 | 99.7 | 46 | 17 |
| Apulia 15 | 1,263 | 1,406 | 2.23 | 1 | 5 | 5.56 | 2 | 15 | 99.5 | 30 | 16 |

Figure 5 refers to the exemplifying case of the Big Town WDN. The panel 5(a) shows the network structure, which retraces clearly the urban structure, while the other figures compare the Empirical density and cumulative distributions of *neighbourhood* nodal degree versus the theoretical Poisson and Pareto distributions. These latter are fitted using the average value of the *neighbourhood* nodal degree and a calibrated value of $\gamma$ of Eq. (5), respectively. As in Figure 4 reports the cumulative distributions using both arithmetical and logarithmic scales. The comparisons clearly show that the Empirical distribution of *neighbourhood* nodal degree is very similar to the Poisson distribution, while it substantially differs from the Pareto one.

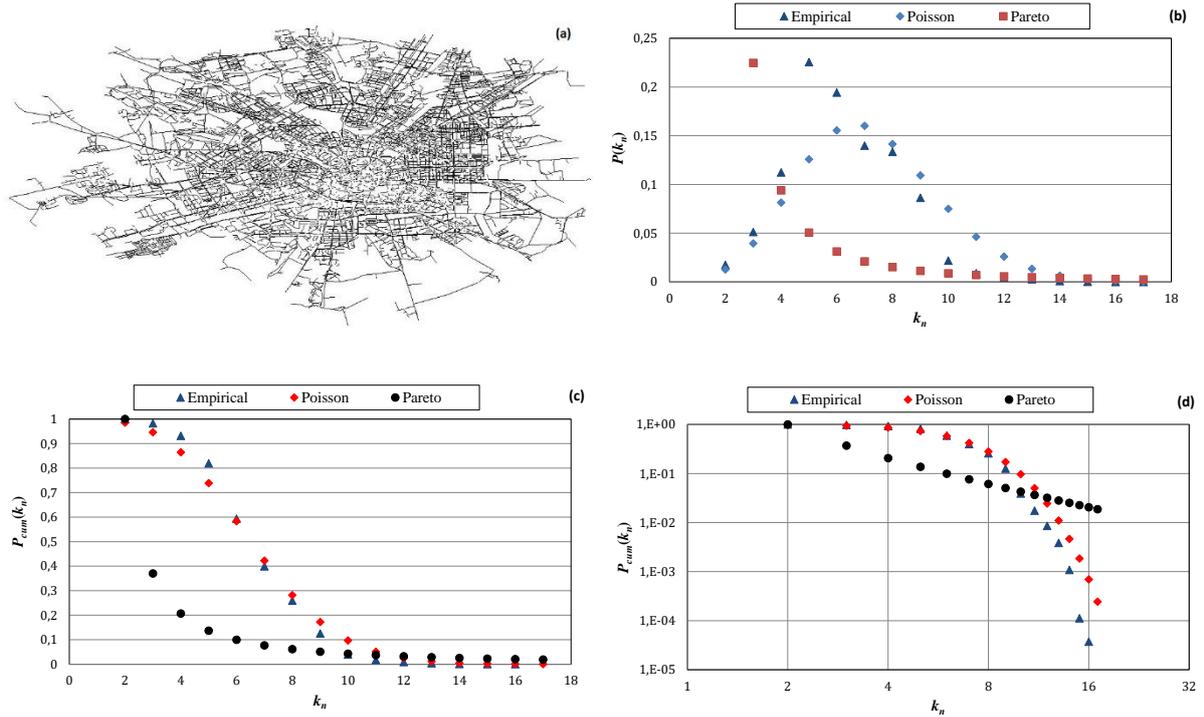

Figure 5. Layout of the Big Town network and Empirical versus Poisson and Pareto distributions.

In the Appendix, the layouts and the cumulative distributions in logarithmic scale for all the WDNs listed in Table I, excluding BBLAWN, Big Town and Exnet, are reported. Apart the case of Exnet network, all the comparisons demonstrate that the Poisson distribution models very well the Empirical distribution of the *neighbourhood* nodal degree. The visual inspection is confirmed by the two-sample Kolmogorov-Smirnov (KS) goodness-of-fit hypothesis test (Massey [18]), determining if the distributions of the values in the Empirical and Poisson samples are drawn from the same underlying population. The results of the test are reported in the third last column of Table I and clearly show a high level of significance. It is worth noting that the Exnet network is the only one having a low significance level (about 26%).

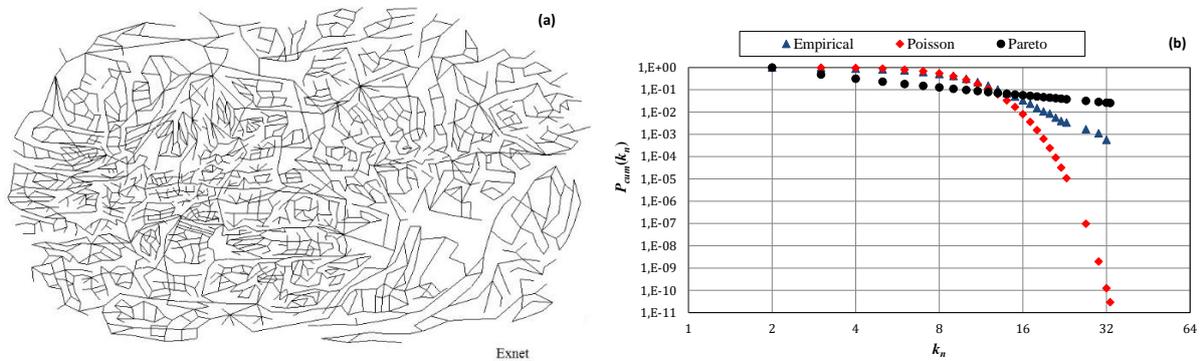

Figure 6. Layout of the Exnet network and Empirical versus Poisson and Pareto cumulative distributions.

Figure 6 shows the Exnet case. It is to observe that Empirical cumulative distribution (in log scale) partially follows the Poisson model up to value of $k_n$ equal to about 13 and then it becomes linear, similarly to the Pareto model. This is the only case, among the analysed network connectivity structures, which shows the features of both Poisson and Pareto models consistently with Boccaletti et al. [4]. Actually, it is to note that (see Table I) Exnet network has the greatest maximum *standard* and *neighbourhood* nodal degree, respectively equal to 11 and 33, revealing in Figure 6 the presence of few hubs.

## 5. DISCUSSION AND CONCLUDING REMARKS

Infrastructure networks are one of most representative cases of spatial networks. The term *spatial* characterizes networks in which nodes are located in a space equipped with a metric (Barthelemy [13]). For most of the infrastructure networks, the space is two-dimensional and the metric is the usual Euclidean distance. Spatial information comes in the network topology typically by the construction process, which is strongly affected by *spatial* constraints. In fact, infrastructure networks are generally manmade and they progressively grow filling the space and balancing connection costs and nodal distances, but also constrained by the impracticality of some connections (Buhl et al. [19]). The existence of such constrains explain why spatial networks generally are not scale free networks (Barthélemy [13] [20]) and, in this line, our findings indicate that the Poisson distribution is generally the best model to describe the network connectivity (as seen by the *neighborhood* nodal degree) of WDNs.

The work by Barthélemy and Flammini [21] about the temporal evolution of urban road networks (see Figure 7) is helpful to understand our finding about WDNs, being these latter networks strongly influenced by the topology of urban patterns. The initial network (see the top-left panel in Figure 7) has a low size and appears quite regular (a ring with some branches is the typical initial configuration). Afterwards, the network evolves and new connections to customer proprieties are built. During this growth phase, the network is affected by the *spatial* constrains but a certain level of randomness emerges also related to the decision of technicians to design considering system redundancy for WDN management; this is due to the different local shapes of such constrains: e.g., different buildings, house blocks, districts, etc. During its evolution and increasing its size, the network can be increasingly classified into the category of random networks (and possibly as small world network). It follows that the Poisson distribution is the best suited to model the network structure of WDNs because the evolution of such systems, although constrained, introduces random characteristics increasing the network size.

The aim of the present work was twofold. Firstly, we proposed the *neighborhood* nodal degree. It is a novel quantity geared to describe the topology in the neighborhood of each node of a network. The second aim has been to show that *neighborhood* nodal degree distribution is suitable to classify infrastructure networks. Differently from the *standard* nodal degree distribution – which generally is uninformative, due to the very limited range of nodal degree values in the infrastructure networks – the *neighborhood* nodal degree exploits the topological information of the nearest neighbors and allows inferring reliable probabilistic models. In particular, we have investigated twenty-two real water distribution networks having different sizes and characteristics. In almost all cases, a Poisson distribution fits the Empirical *neighborhood* nodal degree distribution very well. This result appears in agreement with the characteristics of networks that evolve constrained by urban patterns.

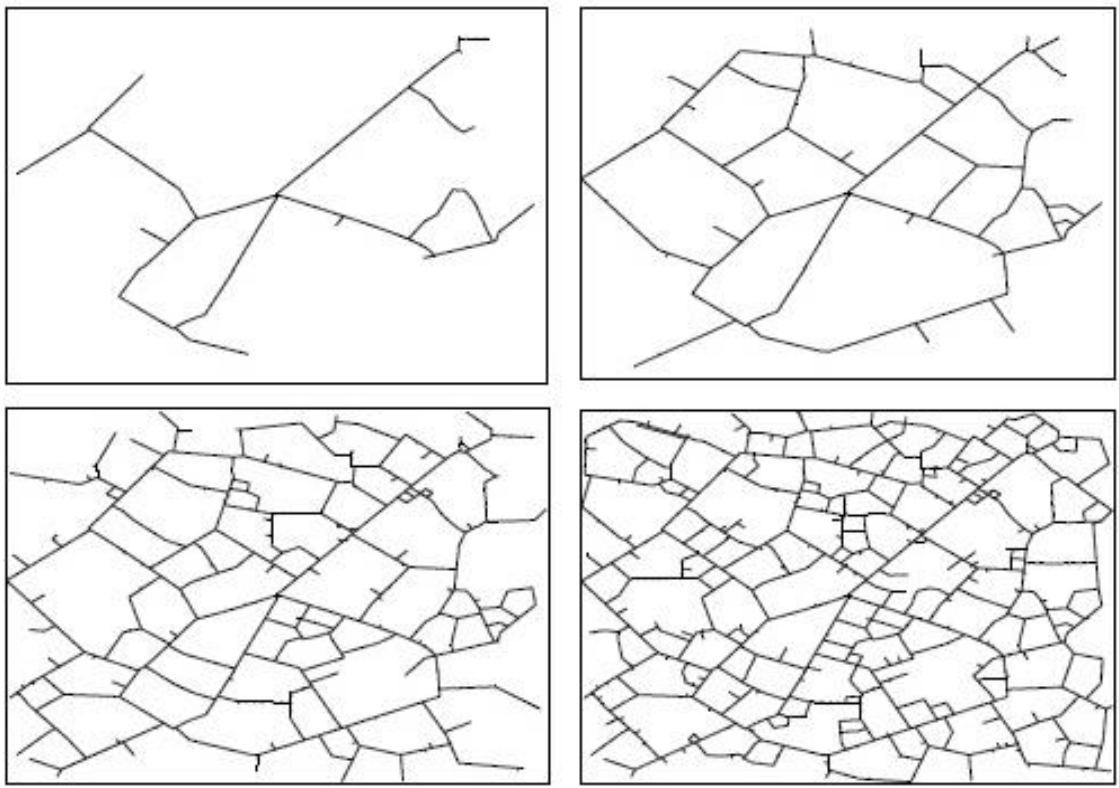

Figure 7. Snapshots of a urban network at different times of its evolution (from Barthélemy and Flammini [21]).

## Appendix

Layout and Empirical cumulative distribution of data versus the theoretical Poisson and Pareto distributions for the nineteen WDNs, which are not reported in the main text.

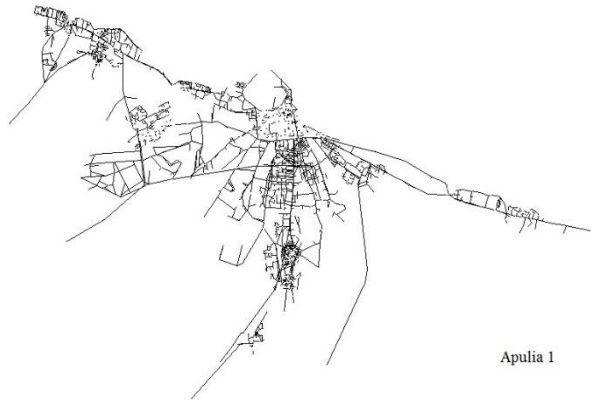
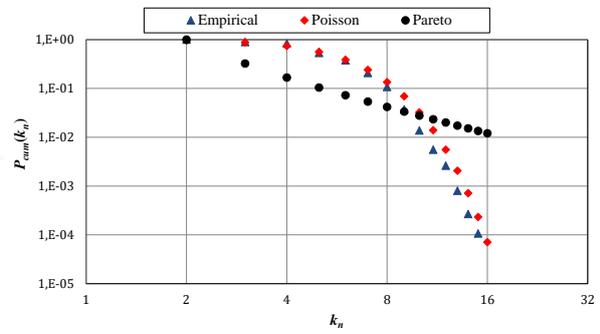

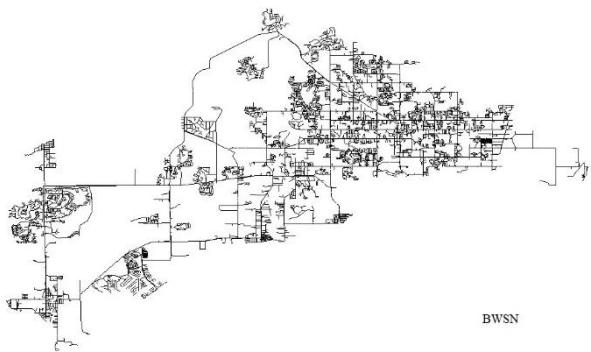
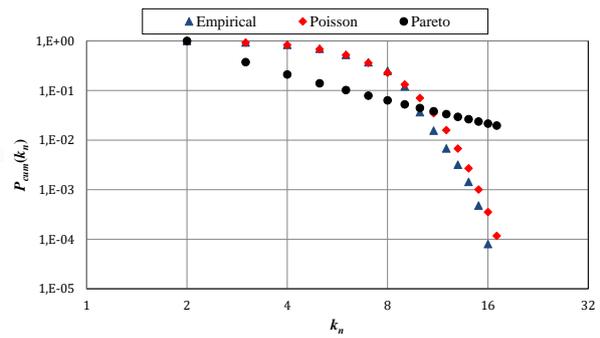

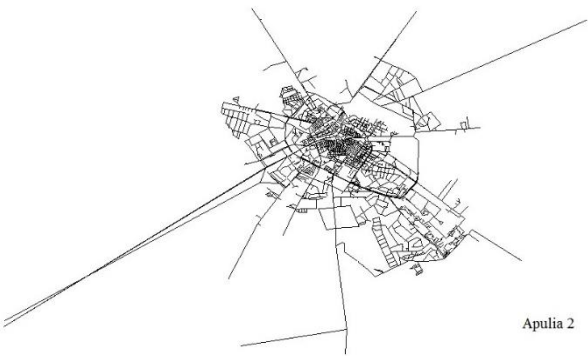
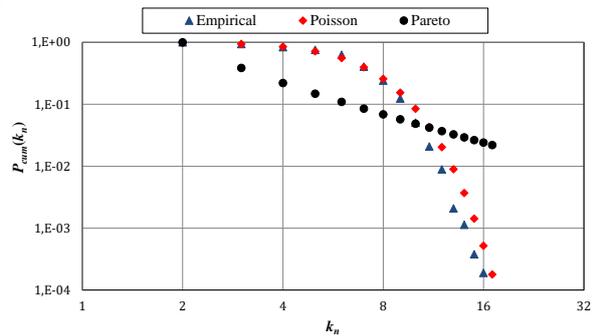

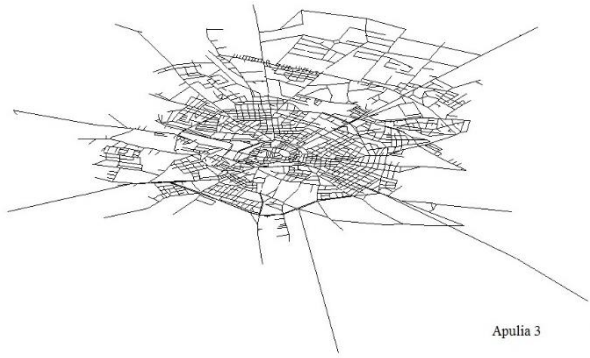
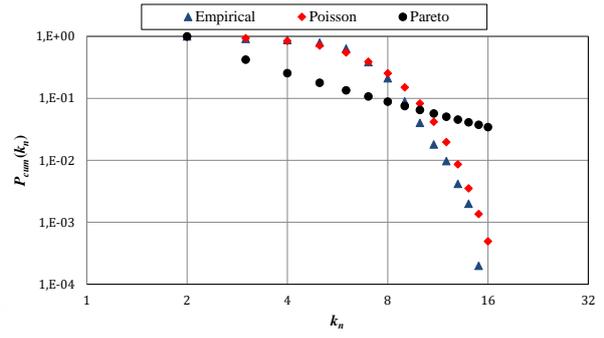

Apulia 3

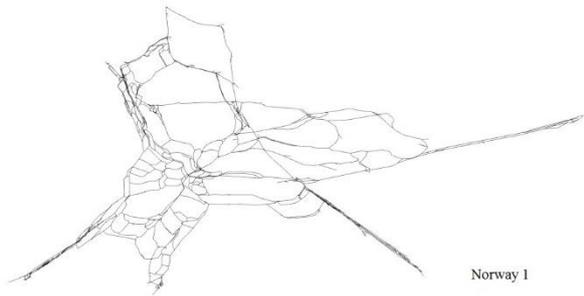
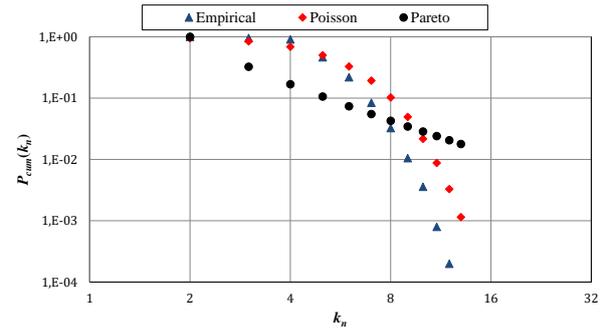

Norway 1

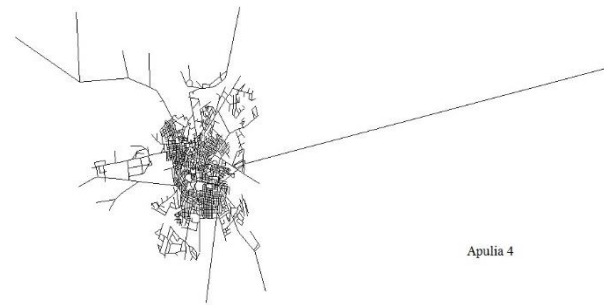
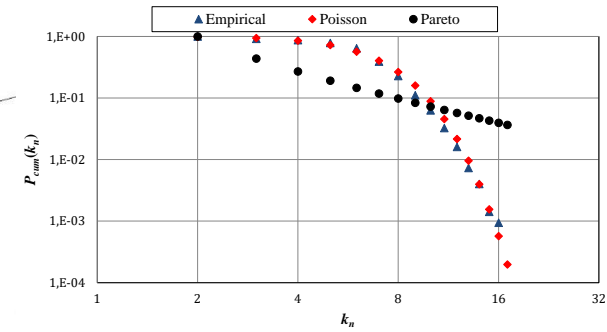

Apulia 4

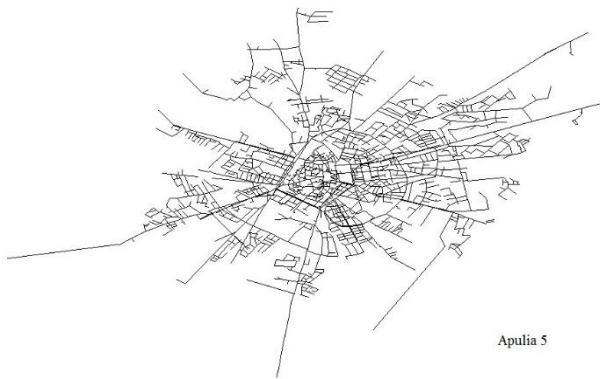
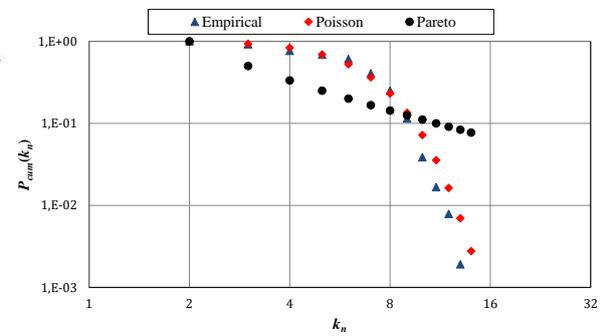

Apulia 5

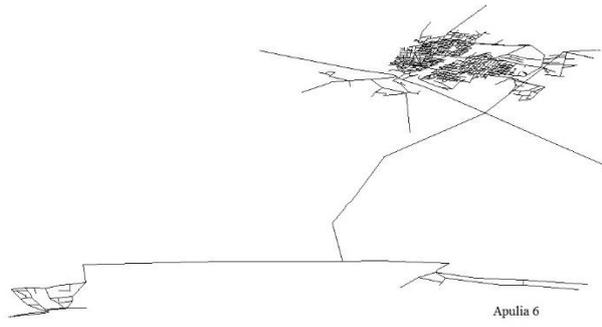
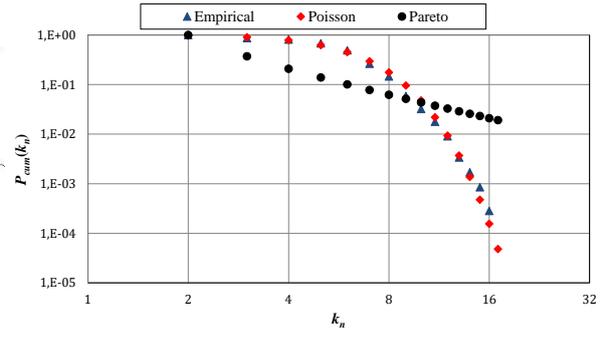

Apulia 6

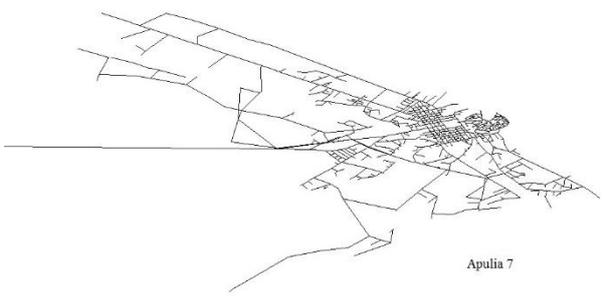
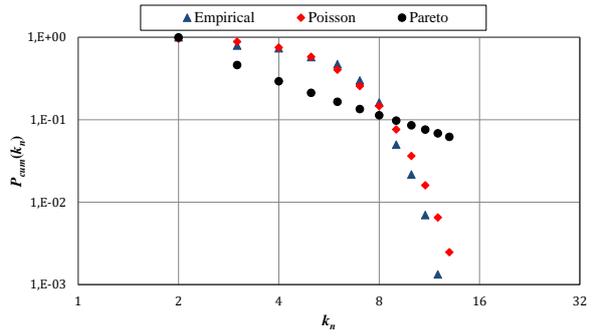

Apulia 7

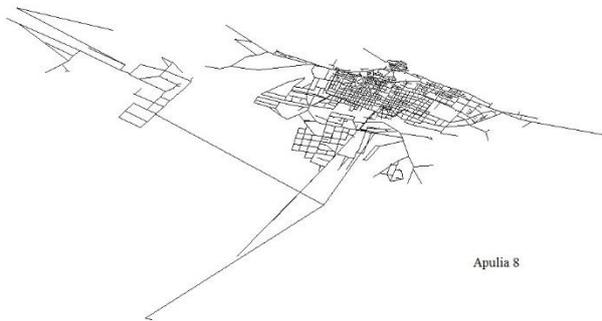
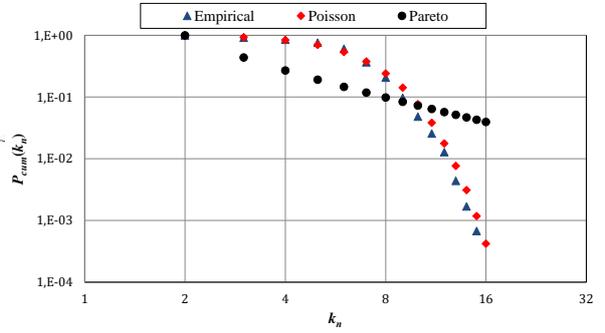

Apulia 8

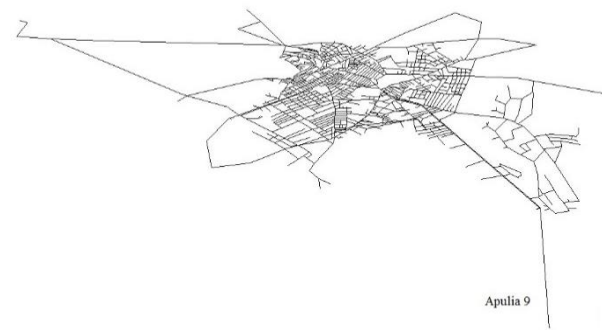
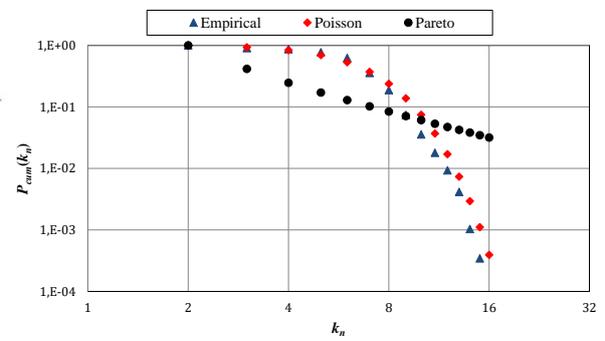

Apulia 9

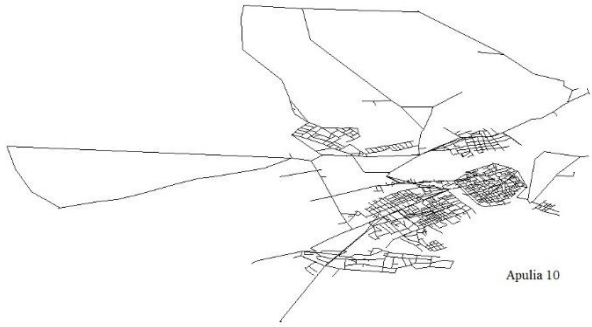
Apulia 10
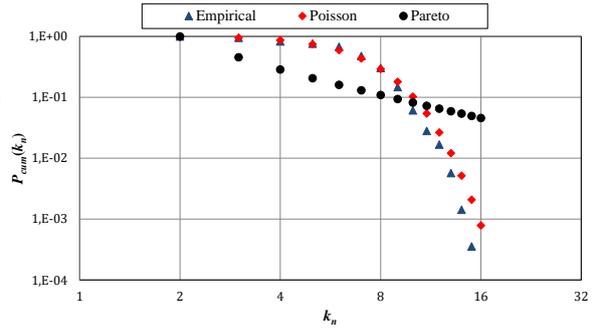

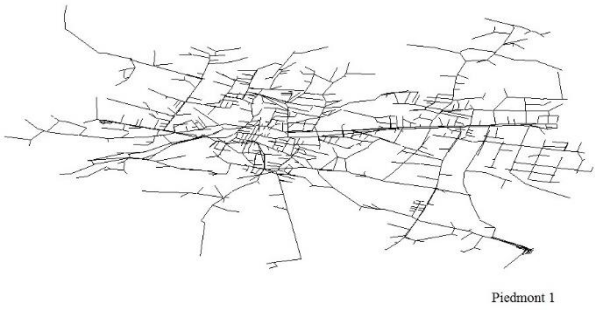
Piedmont 1
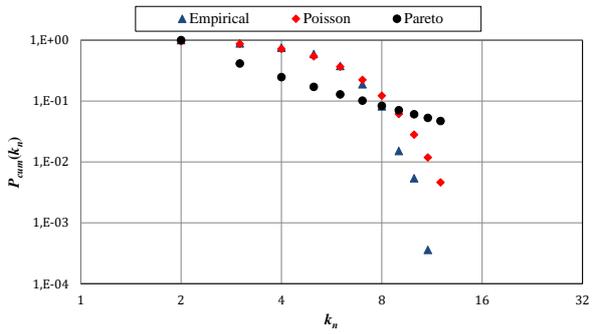

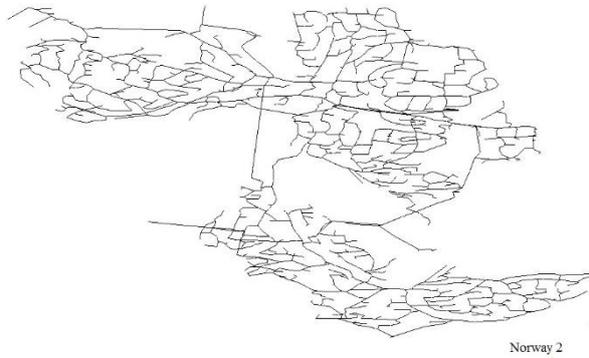
Norway 2
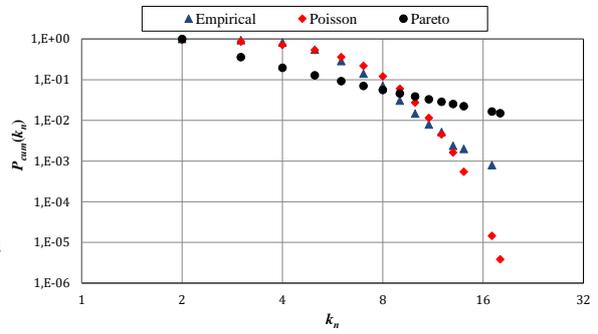

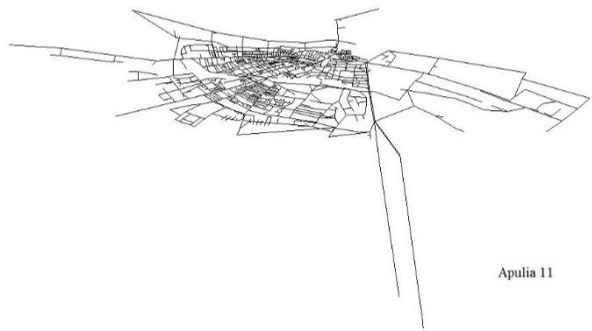
Apulia 11
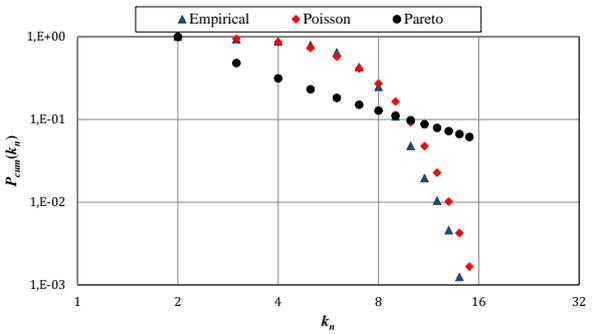

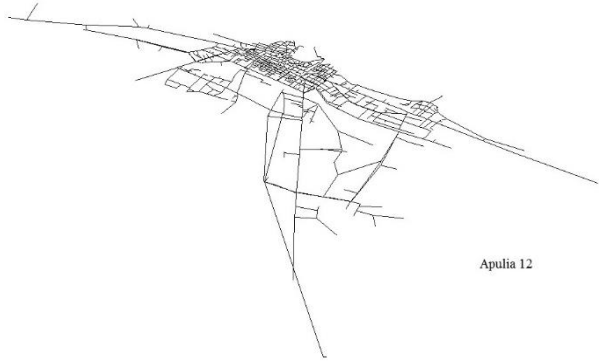
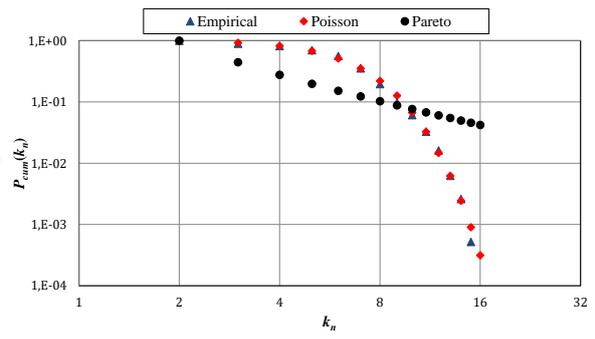

Apulia 12

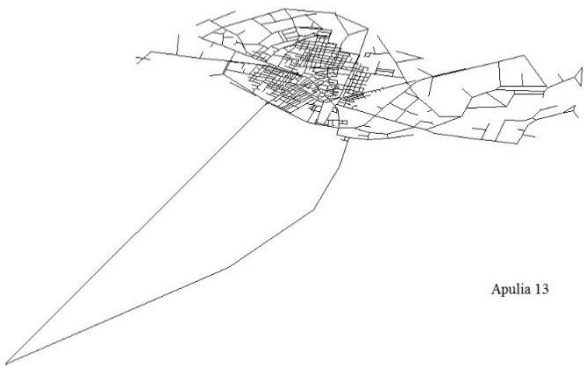
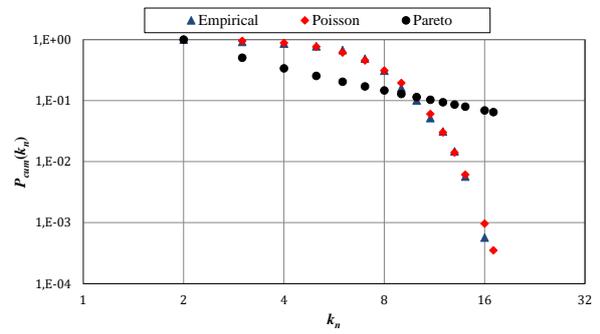

Apulia 13

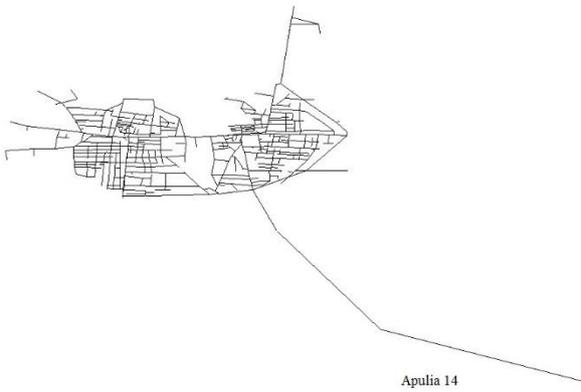
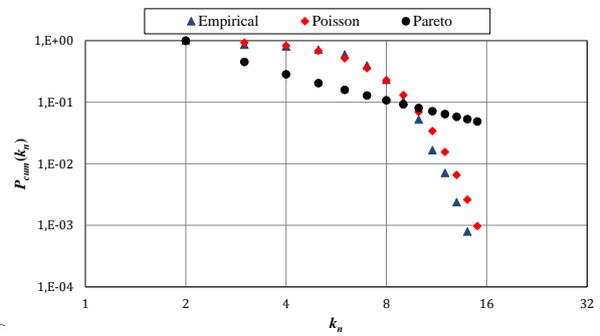

Apulia 14

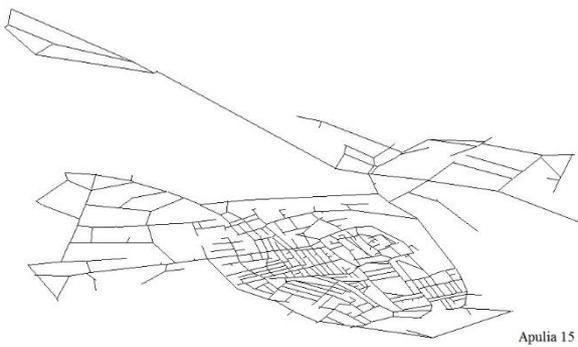
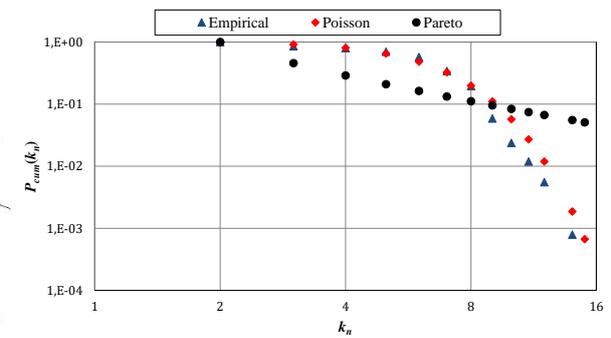

Apulia 15